# Computer Algorithms for Automated Detection and Analysis of Local Ca$^{2+}$ Releases in Spontaneously Beating Cardiac Pacemaker Cells

Alexander V. Maltsev[1],* Sean P. Parsons[2],* Mary S. Kim[1], Kenta Tsutsui[1], Michael D. Stern[1], Edward G Lakatta[1], Victor A. Maltsev[1], Oliver Monfredi[1,3].**

[1]Laboratory of Cardiovascular Science, NIA/NIH, Baltimore, MD, USA, [2]Farncombe Institute, McMaster University, Hamilton, ON, Canada, [3]Division of Cardiovascular Sciences, University of Manchester, Manchester, United Kingdom.

*Short title*: Automated detection and analysis of local Ca$^{2+}$ releases

*These authors contributed equally to this work

**Address for correspondence:

Oliver Monfredi MD PhD

Laboratory of Cardiovascular Science,

Intramural Research Program,

National Institute on Aging, NIH,

251 Bayview Blvd., Baltimore, MD 21224-6825 USA.

Phone: 410-558-8256. Fax: 410-558-8150.

E-mail: oliver.monfredi2@nih.gov

**Abstract:** Local $Ca^{2+}$ Releases (LCRs) are crucial events involved in cardiac pacemaker cell function. However, specific algorithms for automatic LCR detection and analysis have not been developed in live, spontaneously beating pacemaker cells. In the present study we measured LCRs using a high-speed 2D-camera in spontaneously contracting sinoatrial (SA) node cells isolated from rabbit and guinea pig and developed a new algorithm capable of detecting and analyzing the LCRs spatially in two-dimensions, and in time. Our algorithm tracks points along the midline of the contracting cell. It uses these points as a coordinate system for affine transform, producing a transformed image series where the cell does not contract. Action potential-induced $Ca^{2+}$ transients and LCRs were thereafter isolated from recording noise by applying a series of spatial filters. The LCR birth and death events were detected by a differential (frame-to-frame) sensitivity algorithm applied to each pixel (cell location). An LCR was detected when its signal changes sufficiently quickly within a sufficiently large area. The LCR is considered to have died when its amplitude decays substantially, or when it merges into the rising whole cell $Ca^{2+}$ transient. Ultimately, our algorithm provides major LCR parameters such as period, signal mass, duration, and propagation path area. As the LCRs propagate within live cells, the algorithm identifies splitting and merging behaviors, indicating the importance of locally propagating $Ca^{2+}$-induced-$Ca^{2+}$-release for the fate of LCRs and for generating a powerful ensemble $Ca^{2+}$ signal. Thus, our new computer algorithms eliminate motion artifacts and detect 2D local spatiotemporal events from recording noise and global signals. While the algorithms were developed to detect LCRs in sinoatrial nodal cells, they have the potential to be used in other applications in biophysics and cell physiology, for example, to detect $Ca^{2+}$ wavelets (abortive waves), sparks and embers in muscle cells and $Ca^{2+}$ puffs and syntillas in neurons.

**Introduction**

Intracellular localized $Ca^{2+}$ releases from the sarcoplasmic reticulum (SR) via release channels (ryanodine receptors, RyRs) are crucially important events in cardiac cell physiology, as illustrated by the 'local control theory', describing excitation-contraction coupling in cardiac muscle [1]. Individual local $Ca^{2+}$ release events in ventricular myocytes, known as '$Ca^{2+}$ sparks', have been extensively characterized [2]. They are activated by opening of L-type $Ca^{2+}$ channels via $Ca^{2+}$-induced $Ca^{2+}$ release (CICR). Such sparks exhibit stereotyped spatiotemporal behavior, with a local rise followed by decay over a defined time course, having a duration of about 20 ms, and a size of about 2 µm. Traditionally, spark characteristics have been measured by confocal scan line techniques, and algorithms for their automatic analysis have been developed and characterized [3, 4]. More recently, sparks have been studied in two-dimensions, necessitating the development of more advanced analysis programs, with the capability for analysis of spark behavior in both the x and y planes [5].

In cardiac pacemaker cells, local $Ca^{2+}$ releases (LCRs) differ from those in ventricular myocytes. They are spontaneous, rhythmical, arising stochastically during diastolic depolarization [6, 7]. They can be also activated by T-type $Ca^{2+}$ channels [8] or L-type $Ca^{2+}$ channels [9], especially the low-voltage activated Cav1.3 isoform [10]. The LCRs interact with membrane bound $Na^+$-$Ca^{2+}$ exchanger (NCX), contributing crucial inward current throughout diastolic depolarization (review [11]). These LCRs are highly heterogeneous, not stereotyped, appearing in the form of either short lived sparks or locally propagating waves, and ending by spontaneously fading and disappearing or by their fusion into the ensuing action potential (AP)-induced whole cell $Ca^{2+}$ transient. Despite their crucial physiological importance, their spatiotemporal structure has not been systematically studied, and no algorithm for the automatic and objective study of their complex characteristics has been devised.

There has been intensive study of LCRs over the last 15 years using confocal line scanning and, more recently, high speed camera recordings [12]. However, substantial progress has been hampered by the lack of availability of tools to rapidly and objectively analyze such complex recordings generated and thereafter yield full, detailed and reproducible characterization of the entire LCR ensemble. In addition to their spatiotemporal complexity, LCRs occur in morphologically highly heterogeneous, spontaneously contracting SA node cells (SANC). These characteristics have proved to be further major obstacles to automated analysis of LCRs. Programs previously developed for spark analysis in ventricular myocytes have proved to be inadequate in addressing the specific challenges of LCR analysis and characterization in SANC.

In this study, we have addressed these challenges by automating LCR analysis in two dimensional movies of $Ca^{2+}$ signals in SANC, through the development of two novel computerized algorithms. First, we developed a new ImageJ plugin that detects local curvature of the SANC, with subsequent affine transformation to render the cell visually stationary. Thereafter, a second novel algorithm that was written in C++ objectively and automatically detects LCRs. Specifically, the second new algorithm enables complex spatiotemporal characterization of LCRs, including details of LCR birth, death, splitting, merger, and area path travelled within the cell, in addition to more established characteristics like size, intensity and duration. Such detailed characterization is crucial to give novel insights into the fine structure of the LCR ensemble and biophysical mechanisms of pacemaker cell function.

**Materials and Methods**

*Ethics Statement*

The present study conformed to the Guide for the Care and Use of Laboratory Animals, published by the US National Institutes of Health. The experimental protocols were approved by the Animal Care and Use Committee of the National Institutes of Health (protocol # 034-LCS-2019). New Zealand White rabbits (Charles River Laboratories, USA) weighing 2.8–3.2 Kg were deeply anesthetized with sodium pentobarbital (50–90 mg/kg) injected to the central ear vein. The adequacy of anesthesia was monitored in rabbits until reflexes to ear and tale pinch and jaw tone were lost. The guinea pigs (Charles River Laboratories, USA) weighed 500 – 650 grams and were acutely anesthetized with pentobarbital sodium using IP injection at approximately 150 mg/kg doses to an absence of toe pinch reflex and eye membrane retraction.

*SANC Preparations*

Spontaneously beating SANC were isolated from guinea pig and rabbits using essentially the same procedure as previously described [13]. The heart was removed quickly and placed in the solution containing (in mM): 130 NaCl, 24 $NaHCO_3$, 1.2 $NaH_2PO_4$, 1.0 $MgCl_2$, 1.8 $CaCl_2$, 4.0 KCl, 5.6 glucose equilibrated with 95% $O_2$/5% $CO_2$ (pH 7.4 at 35.5°C). The sinoatrial node (SA) region was cut into small strips (~1.0 mm wide) perpendicular to the crista terminalis (CT) and excised. The final SA node preparation consisted of SA node strips attached to the small portion of crista terminalis. The SA node preparation was washed twice in $Ca^{2+}$-free solution containing (in mM): 140 NaCl, 5.4 KCl, 0.5 MgCl2, 0.33 $NaH_2PO_4$, 5 HEPES, 5.5 glucose, (pH=6.9) and incubated at 35.5°C for 30 min in the same solution with addition of elastase type IIA (0.6 mg/ml; Sigma, Chemical Co.), collagenase type 2 (0.6 mg/ml; Worthington, NJ, USA) and 0.1% bovine serum albumin (Sigma, Chemical Co.). The SA node preparation was washed in modified Kraftbruhe (KB) solution, containing (in mM): 70 potassium glutamate, 30 KCl, 10 $KH_2PO_4$, 1 $MgCl_2$, 20 taurine, 10 glucose, 0.3 EGTA, and 10 HEPES (titrated to pH 7.4 with KOH), and kept at 4°C for 1h in KB solution containing 50 mg/ml polyvinylpyrrolidone (PVP, Sigma, Chemical Co.). Finally, cells were dispersed from the SA node preparation by gentle pipetting in the KB solution and stored at 4°C.

### *2D $Ca^{2+}$ dynamics measurements in single SANC*

$Ca^{2+}$ dynamics within isolated single SANC were measured by 2D imaging of fluorescence emitted by the $Ca^{2+}$ indicator fluo-4 using a high-speed Hamamatsu C9100-12 CCD camera with an 8.192 mm square sensor of 512 x 512 pixels resolution or a PCO.edge 4.2 CMOS camera with an 13.2 mm square sensor of 2048x2048 pixels resolution.

To resolve LCR dynamics we acquired images at a rate of 100 frames/second that was possible only using a part of the sensor (e.g. 160 x 512 pixels for Hamamatsu camera or 1280x1280 for PCO camera). In some recording with the PCO.edge 4.2 CMOS camera we used 2x2 binning. Both cameras were mounted on Zeiss Axiovert 100 inverted microscopes (Carl Zeiss, Inc., Germany) with a x63 oil immersion lens and a fluorescence excitation light source CoolLED pE-300-W (CoolLED Ltd. Andover, UK).

Fluo-4 fluorescence excitation (470/40 nm) and emission light collection (525/50 nm) were performed using the Zeiss filter set 38 HE. Cells were loaded with 1.5 µM fluo-4AM (Sigma-Aldrich) for 10 minutes at room temperature. Fluo-4AM was subsequently washed out of the chamber, and $Ca^{2+}$ signals were measured within the ensuing 60 minutes at 35°C ± 0.1°C.

To avoid phototoxicity, fluo-4 was excited only for short periods of time (<5 s). Data acquisition was performed using SimplePCI (Hamamatsu Corporation, Japan) or PCO camware 64 (PCO AG, Germany). In some experiments (n=10 cells, S1 Movie) we also recorded simultaneous membrane potentials by using perforated patch clamp as previously described [12].

***Bathing solution and temperature control***

All $Ca^{2+}$ and AP measurements were performed at 35°C+/-0.1°C (500 µl chamber volume). Temperature was controlled by an Analog TC2BIP 2/3Ch bipolar temperature controller from CellMicroControls (USA). This heated both the glass bottom of the perfusion chamber and the solution entering the chamber (via a pre-heater). The physiological (bathing) solution contained (in mM): NaCl 140; KCl 5.4; $MgCl_2$ 2; HEPES 5; $CaCl_2$ 1.8; pH 7.3 (adjusted with NaOH).

***Software development***

The SANC Analyser tracking contractile motion of the SANC was written in Java as a plugin for the ImageJ image analysis platform (NIH, Bethesda, MD), using the Netbeans IDE. The SANC Analyser plugin with its full source code and pdf user guide and example of image series can be downloaded from http://scepticalphysiologist.com/code/code.html or alternatively from https://drive.google.com/open?id=0BxvuqGAQdt-YdWFDb2Ywd0ZoRXc.

The XYTEventDetector was written in C++ using Microsoft Visual Studio 2010. Additional libraries used in development were libtiff and OpenGL.

The XYTEventDetector is an open source program that can be downloaded from https://drive.google.com/open?id=0B68Q9S0HOGLHSUpBZWp4dEN5RXM

The website provides the following materials:

1) Our program's C++ code with all libraries (needed for program compilation in Visual Studio).

2) msi file for program installation in computers with Windows operating systems.

3) A detailed User Guide that describes software installation and program use to detect LCRs.

4) Data samples for testing and illustration purposes.

**Results**

*LCRs in 2D movies*

Rhythmical stochastic LCR signals in spontaneously beating SANC emerge between AP-induced whole cell $Ca^{2+}$ transients (see example in S1 Movie). We show the simultaneous patch clamp/2D-$Ca^{2+}$ signal data in order to emphasize the importance of having a fast, reproducible and accurate program for the analysis of LCRs in 2D so that their biophysical behavior can be linked to the behavior of the membrane potential and membrane ion channels.

*SANC Analyser*

Tracking $Ca^{2+}$ signals in a particular cell location is problematic when the cell moves, e.g. due to its contraction. Such movement leads to a so-called 'motion artifact', because each fixed pixel in a stack of images does not exactly report the same location or cell neighborhood both within the same beat, and from beat-to-beat in recordings of several consecutive beats.

To solve this issue, we developed a plug in for ImageJ, 'SANC Analyser', which artificially rendered the complex contracting cell stationary. This allowed identical geographical locations to be followed accurately on a within beat and between beats basis. An example of such analysis and image transformations is illustrated in S2 Movie (a contracting cell before analysis) and S3 Movie (stationary cell after transformation).

Our SANC Analyser is a set of algorithms that tracks the contractile motion of the SANC, and using the tracked points, affine transforms the image series so that the cell appears motionless. Tracking is based on the hypothesis that the local curvature of the cell's long axis is conserved under deformation/contraction, i.e. a bend in the cell is a fixed point which can be tracked during contraction.

A "spine" along the cell's long axis is determined by the centre of mass. With the cell approximately parallel to the *x* axis of the image series *i(x,y,t)*, the centre of mass ($\zeta$) along the *y* axis is:

$$\zeta(x,t) = \sum_y y a(x,y,t)^{1.5} / \sum_y a(x,y,t)^{1.5} \tag{1}$$

$$a(x,y,t) = \begin{cases} i(x,y,t) - b & i(x,y,t) > b + 3\sigma \\ 0 & \text{otherwise} \end{cases} \tag{2}$$

where *b* and $\sigma$ are the mean and standard deviation of the background fluorescence. $\zeta(x,t)$ are the *y* coordinates of the spine (Fig 1A), for which the local curvature (*k*) is,

$$k(t,x) = \zeta_{xx}(x,t) / (1 + \zeta_x(x,t)^2)^{3/2} \tag{3}$$

The partial derivatives of $\zeta(x,t)$ are calculated by Savitsky-Golay filters [14].

As the spine is the centre of mass, its local curvature is a function of: 1) the overall curvature of the cell; 2) bumps in the cell's silhouette which are asymmetric with respect to the long axis and similar asymmetries in fluorescence intensity. The first two (which are really the same thing at different scales) should be conserved during contraction and are therefore a basis for tracking. Asymmetries in fluorescence intensity from LCRs will interfere with this if they are large enough. This can be seen in $k(t,x)$, the local curvature "map" (Fig 1B). Horizontal bands of constant curvature are seen that bend when the cell contracts. Just prior to each contraction these bands become a little noisy due to large LCRs which precede and merge into the AP. This noise means that the bands cannot be tracked automatically by a simple least difference algorithm, for example. Instead lines of constant curvature are tracked semi-manually, as follows.

A single contraction is identified manually with a rectangular ROI. Contractions in SANC are very stereotyped, so the remaining contractions are identified automatically by a least difference algorithm in comparison with the manually identified contraction (Fig 1B, blue lines). A horizontal "guide" line is made along a band of constant curvature or its border. Within one identified contraction the guide is bent manually to match the bend of the curvature band by making a polyline selection/ROI. This bend is then iterated automatically for the other contractions, again relying on the stereotyped nature of the contractions. The resultant line, $x = L(t)$ (Fig 1B, red lines), corresponds to a point on the spine, $\{x = L(t), y = \zeta(L(t), t)\}$. Multiple lines (points on the spine) are made either by the procedure above or by interpolation between neighboring lines.

A series of points along the spine form the basis of a triangular grid (Fig 1C and Fig 1D). Each pair of points is the base of two isosceles triangles, one below and one above the spine, of user defined height. Each pair of neighboring triangles define an intervening triangle of opposite orientation. The grid is used to affine transform the series. Where the three points of a triangle form the matrix:

$$\mathbf{M} = \begin{bmatrix} x_1 & x_2 & x_3 \\ y_1 & y_2 & y_3 \\ 1 & 1 & 1 \end{bmatrix} \quad (4)$$

the transform from {x, y} in a reference image (to which all images are to be transformed) to {x', y'} in an image to be transformed is:

$$\begin{bmatrix} x' \\ y' \\ 1 \end{bmatrix} = \mathbf{M_d} \mathbf{M_r}^{-1} \begin{bmatrix} x \\ y \\ 1 \end{bmatrix} \quad (5)$$

where $\mathbf{M_r}$ is the reference triangle and $\mathbf{M_d}$ the transforming triangle.

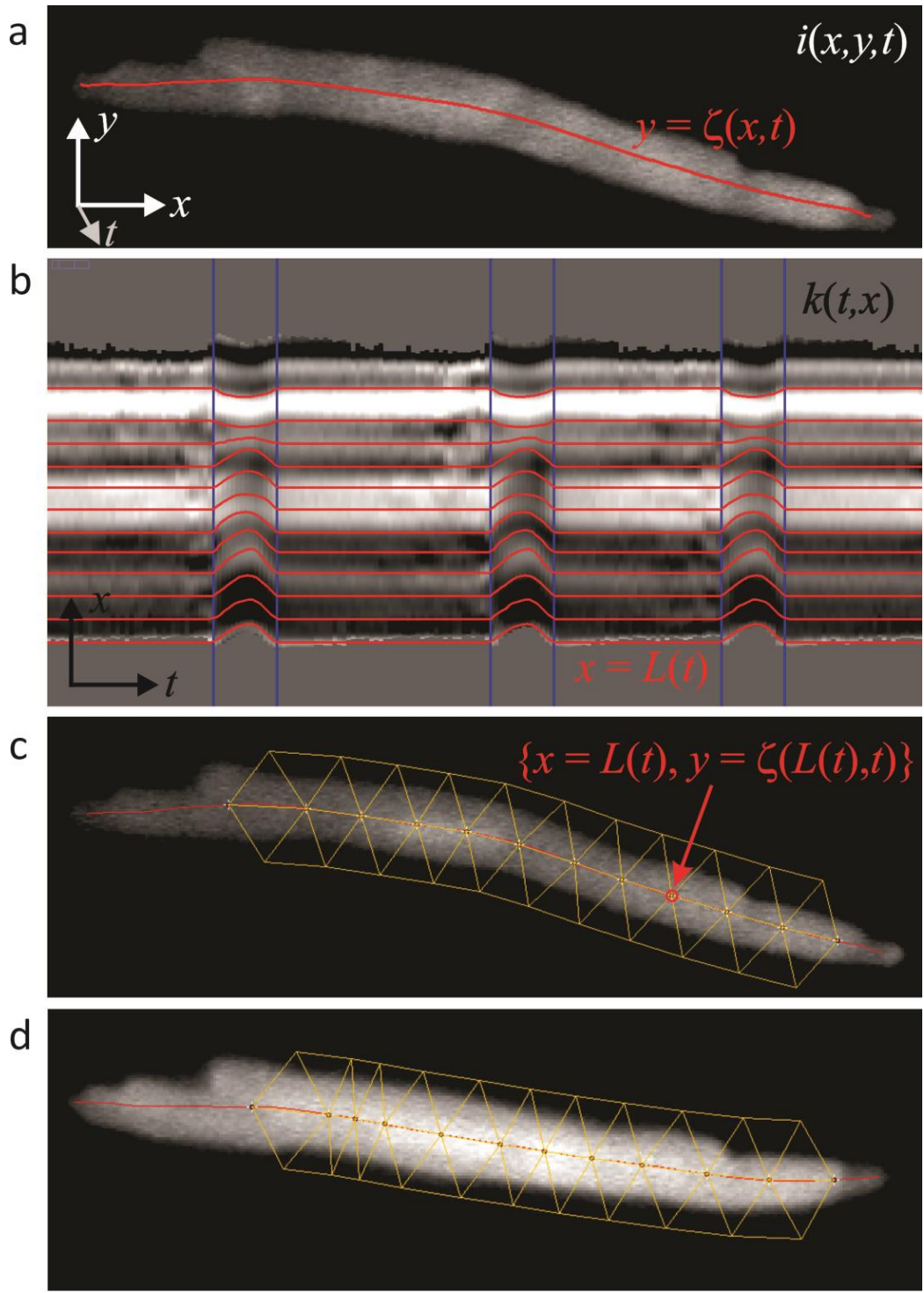

**Fig 1. A novel algorithm ('SANC Analyser') to render spontaneously beating SANC stationary.**
a: the algorithm relies on the generation of a center of mass, or COM, line. b: the COM line is used to produce a map of local curvature, the lines on which are tracked semi-automatically across consecutive contractions. c: a triangular grid is generated on top of the cell, and this is used to affine transform the series (d). Extensive details are given in the text.

*Notes for practical use of SANC Analyser*

The automatic identification of contractions does require some degree of care and experimentation in terms of selecting the correct kind of contraction to make the original archetypal contraction when you hit the 'set contraction' button. For example, if too bright a contraction is selected with 'set contraction', then 'match contraction' will not find subsequent contractions if there is bleaching over the time period of the recording. In such circumstances, choosing a contraction with fluorescence intensity somewhere in between the maximum and minimum intensities associated with all of the contractions over a recorded time will most likely cause the program to accurately find all contractions. Experimentation with this is required by the user until experience is gained. Similarly, identification of contractions can be improved by altering the 'match threshold' number – this becomes more sensitive when higher numbers are inputted, and the authors have found that increasing the default of 20 to 30 often yields satisfactory results in terms of picking up contractions that were not picked up when the 'match threshold' was set at 20.

This phase of the analysis is very important, and strenuous efforts should be made to identify as many contractions as possible. If a substantial number of contractions cannot be automatically identified as detailed above, then the results from that particular cell should be discarded. For further detail please see our user program guide that is provided online together with the program code and plugin (see Methods).

*LCR Detection*

Our algorithm was implemented within an open-source computer program XYTEventDetector (see Methods). The user Interface for this program is shown in Figure 2 and a detailed user guide is provided online together with the program code and program installation file. Below we provide principles of LCR detection together with a short description of specific program interface parameters customizing the execution of these principles.

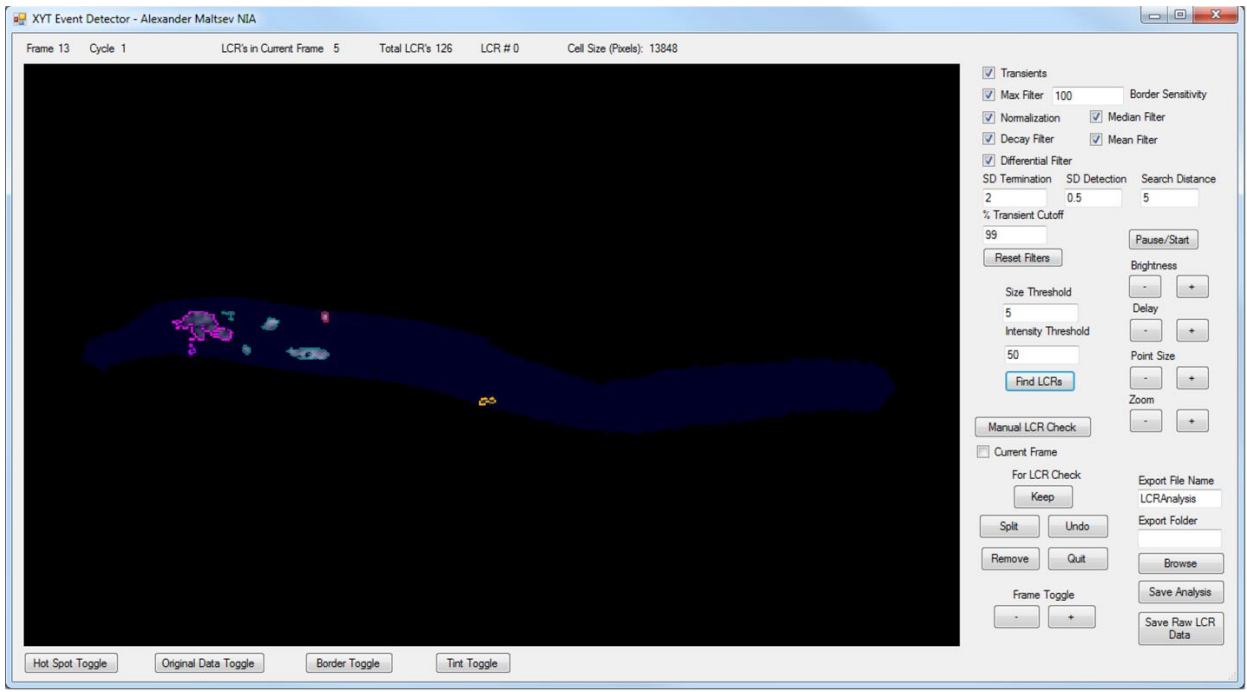

**Fig 2. User interface of** t**he XYTEventDetector.**
The line above the main window provides information on current frame number, cycle number, the number of LCRs in the current frame, the total number of LCRs found, and cell size in pixels. Main window shows progress of LCR analysis in a given cell. When LCRs are found, they are outlined by different colors. Buttons underneath the main window toggle on/off LCR border, cell area (Tint Toggle), original data, and Hot spots, i.e. locations of predominant LCR occurrences. 'User-definable' options are to the right of the main window. "Transients" checkbox informs program that data has transients. If this box is unchecked programs analyses LCR activity regardless to pacemaker cycle, e.g. in permeabilized cells. The filters and their settings are uppermost, while modifiable detection characteristics for LCRs are below the filter settings. Checking each box executes the respective function. The settings for verification of LCRs are below this, with the ability to inspect and edit individual LCRs in each frame. Exporting options are given at the lower right of the analysis pane.

The sequence of LCR detection is illustrated in Figure 3. Our algorithm isolates the cell cytoplasm from the background, and subsequently normalizes the cytoplasmic signals, eliminating the global signal of the AP-induced whole cell $Ca^{2+}$ transient. It then decreases the remaining recording noise using spatial filters that the user defines for given experimental conditions and recording noise, which will vary from cell-to-cell. Thereafter, any local signal that remains is considered a possible, or 'candidate', LCR. These possible LCRs are then further refined with threshold algorithms for signal amplitude and spatial size.

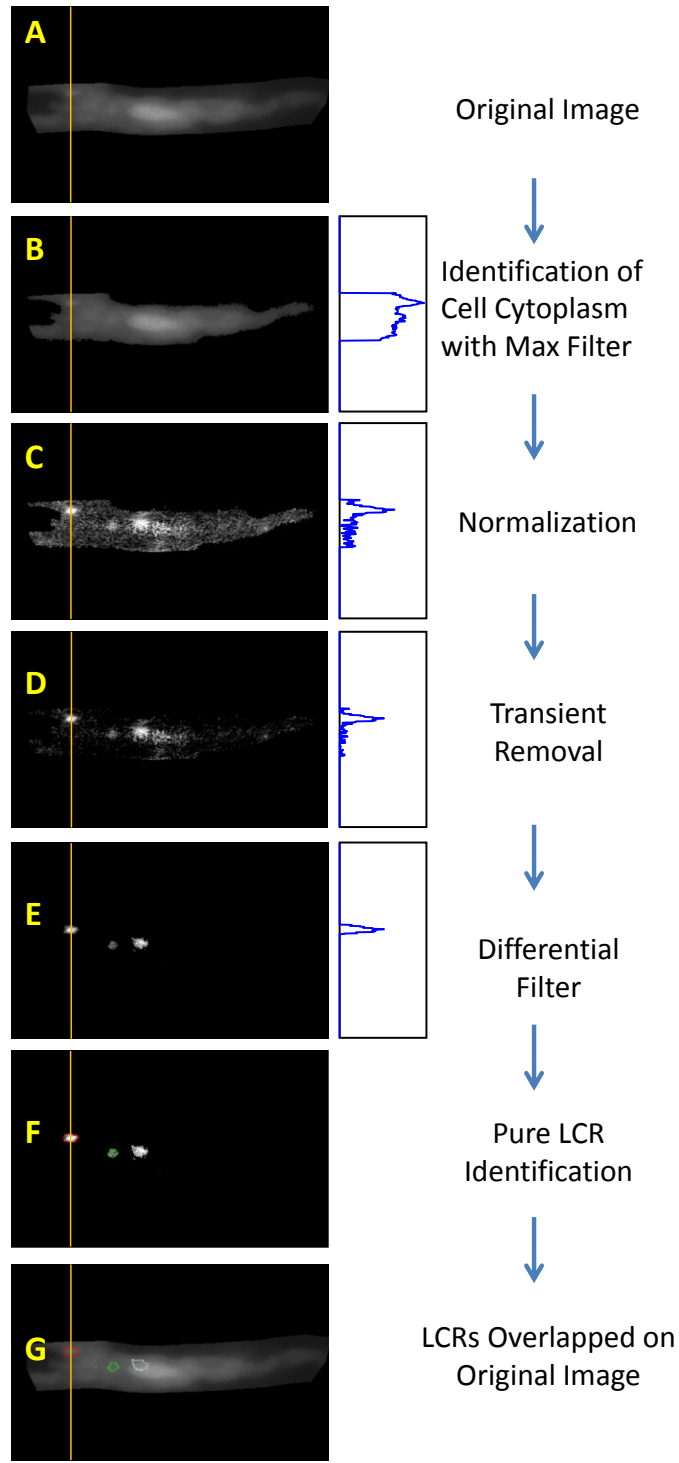

**Fig 3. The sequence of image processing to detect LCRs.**
The evolution of a representative LCR profile along Y axes (average Fluo 4 signal within 1 μm band) is shown in blue plots. The LCR signal becomes clearly separated, from the noise and background. Images in panels C,D, and E were contrasted to visualize the detected LCRs.

To identify the pixels representing the cell cytoplasm, the program uses a custom filter, which we termed the 'Max Filter' (Fig 2, Fig 3A, Fig 3B). It calculates the maximum value of each pixel as a function of time. The background is much darker than the cell itself, thus taking the mean of all the maximum values gives a cut point for isolating the cell cytoplasm. Pixels with maximum values greater than the cut point are considered part of the cell area and the rest as part of the background.

Once we have identified the pixels of cell cytoplasm F(x, y, t), we go on to normalize their signal (Fig 3C), taking 255 as the number of levels of gray. It is executed by the "Normalization" box in the program interface. This is the first step in separating LCRs from noise:

$$F_{norm} = 255*(F-F_{min})/(F_{max} - F_{min}) \qquad (6)$$

If there are AP-induced Ca transients in the data, it is important to separate the true local rise of the LCR signal from the global transient. This is accomplished by the "Decay Filter" in the program interface. We take the global transient as the average signal over the entire cell, $F_a(t)$ at each time frame. Then we subtract $F_a(t)$ from each pixel $F_{norm}(x,y,t)$ value inside the cell area over the entire time series, and any pixel value that becomes negative is clipped to be 0 (Fig 3D):

$$F(x,y,t)_{no\ transient} = F_{norm}(x,y,t) - F_a(t) \qquad (7)$$

Then to identify cell locations in which the signal changes over time, we apply a custom differential filter (Fig 3E, "Differential Filter" box in user interface). This is a composite spatiotemporal filter that integrates the signal spatially in all frames and then calculates frame-to-frame differences in each pixel. The spatial averaging width is defined by interface parameter "Search Distance", that is the distance around each pixel used for averaging. By default, Search Distance is set to 3 and defines a grid area of 7x7 pixels with the tested pixel in the middle of this grid. This default setting was used in majority of our LCR detection analysis with the Hamamatsu C9100-12 CCD camera, defining the averaging area of $7*7*0.2539^2 = 3.16$ $\mu m^2$. Thus, this complex filter excludes pixel-to-pixel noise and it is also very effective in excluding locations with constant fluorescence due to heterogeneous indicator loading within the cell cytoplasm and organelles such as nuclei. This identifies locations of potential LCR-related $Ca^{2+}$ dynamics and it is important for subsequent detection of LCR birth/death events.

The "Differential Filter" performs further important discriminations based on signal strength, spatial distribution, and its timing with respect to the beginning of the AP-induced Ca transient. This function can be customized by tuning parameters described below for specific recordings. We first renormalize the pixels using equations 6 and 7. For a pixel to be considered as a part of a newborn or existing LCR, the signals must surpass a certain threshold given by the parameter "SD Detection" in the user interface, which is set by the user as a factor of the standard deviation of the signal in the cell. By default, this parameter is set at one standard deviation. We also assume that if a pixel is inside a possible LCR then it is also part of the LCR. Thus, to account for pixel anomalies inside LCRs, we have assign LCR candidacy to a pixel if 7 neighbors around it have LCR candidacy. Additional discrimination is applied to LCRs which merge to the AP-induced transient. The program parameter "% Transient Cutoff" defines the level of the onrushing transient at which the program stops looking for LCRs, and accepts that the $Ca^{2+}$ transient has overwhelmed and engulfed the cytoplasm beyond this point to the extent that LCRs can no longer be detected or discriminated. Finally, the "Differential Filter" determines whether the pixel intensity drops low enough to be considered no a part of an LCR. The corresponding input parameter is "SD Termination".

We have also two additional (optional) filters, a median and a mean filter, to get rid of any remaining noise, depending on the presence of 'salt-and-pepper noise' (sparsely occurring white and black pixels, also known as 'data drop-out noise', 'intensity spikes' or 'speckle').

So far by using different types of filtering, we effectively reduced noise and identified locations potentially linked to LCR activity, i.e. birth and death. The next step is to identify the new LCRs and to define the dynamics of existing LCRs (Fig 3F, Fig 3G). This procedure starts with the "Find LCRs" button in the interface. In addition to LCR detection sensitivity parameters that exist within the "Differential Filter" described above, there are 2 additional sensitivity parameters to further fine tune LCR detection for optimal performance in rejecting noisy, smaller LCRs or to tailor the program to the specific aims of a particular study (for example, to detect selectively only large LCRs).

The program searches for clusters, i.e. sets of pixels having common borders, which would satisfy specific conditions as described below. A newborn LCR, a cluster of pixels of any morphology, in a certain frame must be composed of a minimum number of pixels, as defined by the parameter "Size Threshold". In practice, this determines the minimum starting size of an LCR in pixels. Increasing the value gets rid of noise but risks failing to detect smaller LCRs. A newborn LCR must be also sufficiently bright (on a scale of arbitrary units 0-255 in our program).  It is important to note that the two sensitivity tests act together with the Boolean function AND, i.e. a newborn LCR must be large and bright enough to pass these tests otherwise it is considered as noise and therefore discarded.

The program subsequently looks further to see if any of these candidates are representative of the dynamics of an existing LCR. We consider that if a candidate (or a cluster of candidates) has a common border with any existing LCR, it becomes a part of that LCRs. Finally, a cluster becomes a newborn LCR if its size and total signal surpass respective user-defined threshold levels. These two sensitivity parameters rejecting small signal fluctuations are given in the program interface as "Size Threshold" and "Intensity Threshold", representing respectively the total number of pixels and the sum of intensities in the smallest LCR.

After a pixel is considered a part of an LCR, it will continue to be part of that LCR until the LCR is presumed terminated, for example by a collision or by merging into the $Ca^{2+}$ transient. In the case of LCR decay, a pixel is considered part of a possible LCR until the average signal in its series becomes less than a factor of the standard deviation that is also by default set at one in the interface parameter "SD Termination" described above.

With dynamics in play (Fig 4, S4 Movie), we assume that when an LCR separates, the separated groups of pixels are still part of the same LCR. When LCRs collide, we consider the largest one as the survivor, which incorporates the other pixels. The other LCR is considered terminated by the collision.

**Birth and death at the same cell location**

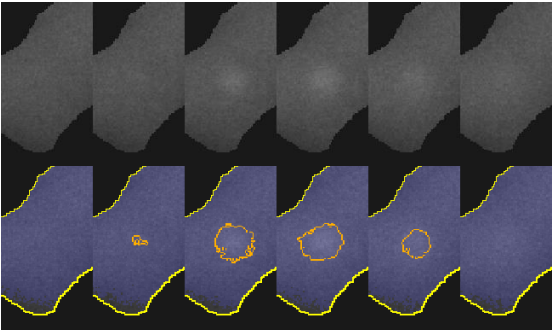

**Progression and propagation**

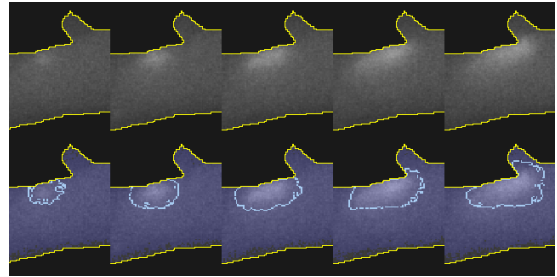

**Collisions/Merges**

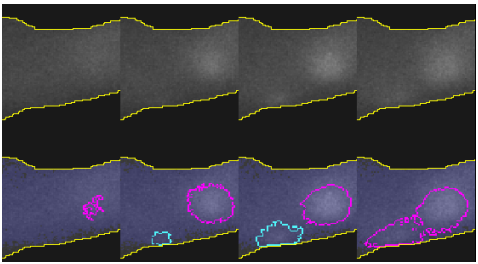

**Separation**

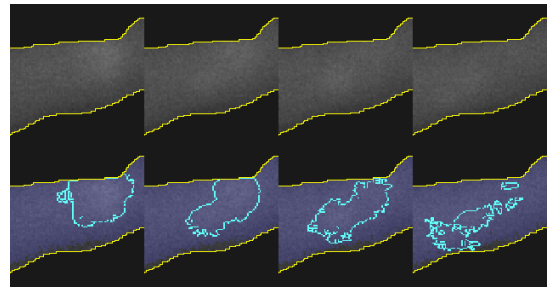

20 μm

Fig 4. Examples of LCR key events, detected by XYTEventDetector including birth, death, propagation, collision, and separation.

*Output LCR parameters*

The parameters outputted by the XYTEventDetector program are as follows:

max_x, min_x, max_y, min_y [pixels]: coordinates of an LCR.

Start Area [pixels]: area of LCR at its birth, i.e. in the first frame in which it is detected.

Max Area [pixels]: maximum area that an LCR reaches during its lifetime.

Path Area [pixels]: the area of the path traveled by the LCR during its lifetime. This characteristic is new: it was not defined for stereotyped $Ca^{2+}$ spark events in ventricular myocytes, but it is especially important in SA node cells because it gives a quantitative characteristic of CICR for every LCR. Specifically, it reflects the greater ability of later occurring LCRs to propagate, self-sustain, and in doing so ultimately have a greater effect on NCX and therefore on diastolic depolarization. This parameter can be used to characterize the impact of individual members of the SANC ensemble. At the same time, this parameter serves a rather reductionist approach, reflecting only release propagation (RyR recruitment to fire) and does not indicate the amount of $Ca^{2+}$ released; therefore, it will be best used together with other parameters to fully characterize the system.

Birth From Nadir [#frames]: discriminates LCRs into categories: a negative value of this parameter indicates that this is an "early" LCR, while a positive value indicates that this is a "late" LCR.

LCR Cycle[#frames]: number of frames from AP-induced transient peak to LCR birth, known also as LCR period. It indicates the timing of local SR "calcium clock" in terms of "coupled clock" theory [15]. Specifically, AP-induced $Ca^{2+}$ release depletes the local SR and resets the refilling to the "empty" state. Once the local SR gets refilled with $Ca^{2+}$ to a threshold level of spontaneous release, it generates an LCR. The time from the reset to LCR occurrence reflects the $Ca^{2+}$ clock function.

Birth From Start [#frames]: number of frames from the beginning of the image stack. It is different from LCR cycle in that it shows LCR birth, independent of AP-induced transient. This parameter is helpful in analysis of, for example, permeabilized cells that do not fire APs.

Duration [#frames]: total # frames for which the LCR lived.

Half Max Amplitude[arb. units]: half of the maximum amplitude of the LCR.

Dead, Dead By Collision, Dead By Transient, Dead By Stochastic Attrition [Boolean]: indicate if LCR dead or alive at the end of the analysis and, if dead, by which mechanism. The program output also generates a separate file with individual LCR traces.

Many of our outputted parameters are similar to those from previous automated spark analysis programs. For example, Steele *et al.*'s xySpark program generated the following parameters [5]: frame number, spark coordinates, amplitude, frequency, full width at half-maximum amplitude (FWHM), $t_{1/2}$ of the descending phase, and $r^2$ value for Gaussian fits. Picht *et al.* similarly generated the following parameters [4, 16]: amplitude, full width at half maximum, full duration at half maximum, full width, full duration and time to peak. Thus, our program outputs LCR/spark characteristics similar to previous programs, with the added benefit of additional parameters that lend even greater insight into LCR behavior, especially related to their local propagation via CICR.

*Sample analysis*

To illustrate the functionality and capabilities of our new method, we have performed a sample analysis in 16 cells, comprising 8 rabbit cells and 8 guinea pig cells, measured by two different cameras. The results of our analysis of key LCR parameters are given in S1 Table and S2 Table. The LCRs are also illustrated in 2 movies with LCR borders marked by different colors, S4 Movie and S5 Movie, for rabbit and guinea pig cells, respectively. The S1 and S2 tables also provide all respective program parameters used to analyze LCRs in each cell, thus providing a clue to parameter selection for best performance.

The program output generates individual LCR traces. In Figure 5 shows an example of LCR traces plotted together with the entire LCR ensemble signal (i.e. a formal sum of all LCR traces) and with the whole-cell $Ca^{2+}$ signal measured in the entire cell area. Small LCRs appear at the beginning of the pacemaker cycle, then individual LCR grow in size, with the concurrently growing LCR ensemble signal (red thick line).

The LCR propagation is captured by the program as seen in the movies (S4 Movie and S5 Movie), and is evident in Figure 5 by the increase in path area towards the cycle end. These movies show that some of the LCRs behave like small waves that travel a variable distance within the cell before either dying out or merging into the oncoming whole cell $Ca^{2+}$ transient, other (typically smaller) LCRs travel very little or not at all.

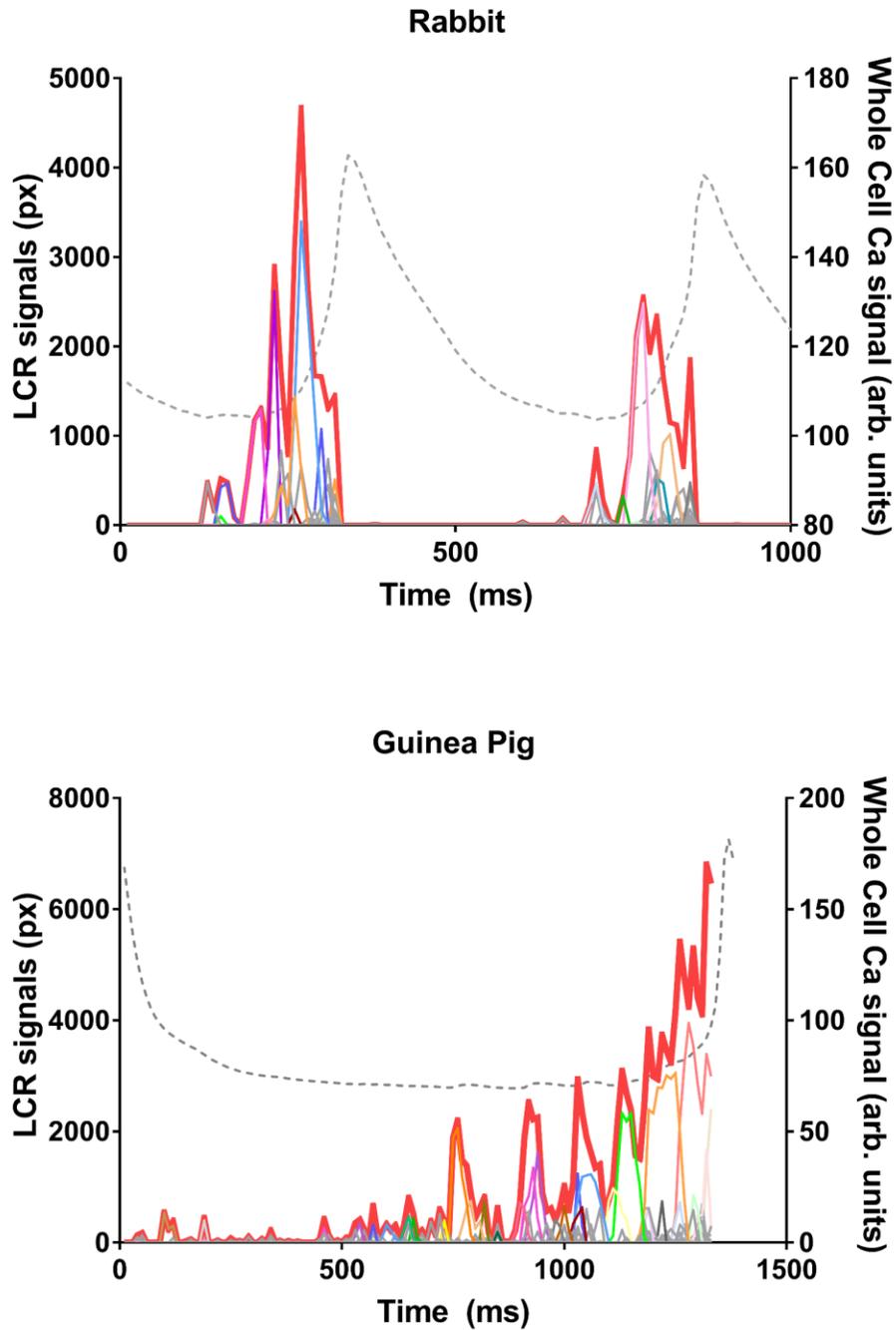

**Fig 5. Individual LCR traces identified by XYTEventDetector.**
Each panel shows examples of individual LCR traces with their integrated ensemble signal during diastole. The LCR traces are shown by fine lines of different colors while the ensemble signal is shown by the thick red line. Traces represent LCR size in pixels as a function of time. Whole cell $Ca^{2+}$ signal is also shown for comparison by dashed line. Upper panel shows data in rabbit and bottom panel shows data in guinea pig SANC. Data were acquired using the PCO camera.

The total number of LCRs per cycle substantially varied in cells of both rabbit and guinea pig (see S1 Table and S2 Table). For example, in rabbit cells these numbers were in the range of 44 to 119, i.e. substantially larger compared to an average of 14 LCRs per cycle previously detected manually in rabbit SANC [12]. This increase was clearly due to detection of the aforementioned 'smaller' LCRs (first bin in histograms in Fig 6).

Also, the LCR duration on average was much shorter, about 13 ms in rabbit, compared to the previous finding of 40 ms. The larger number of automatically detected LCRs and their shorter average duration indicates that the program detects small short-lived events that are hard to detect manually and in an unbiased fashion. Furthermore, LCRs on the decaying transient are detected more faithfully and accurately here, with the transient having been effectively removed.

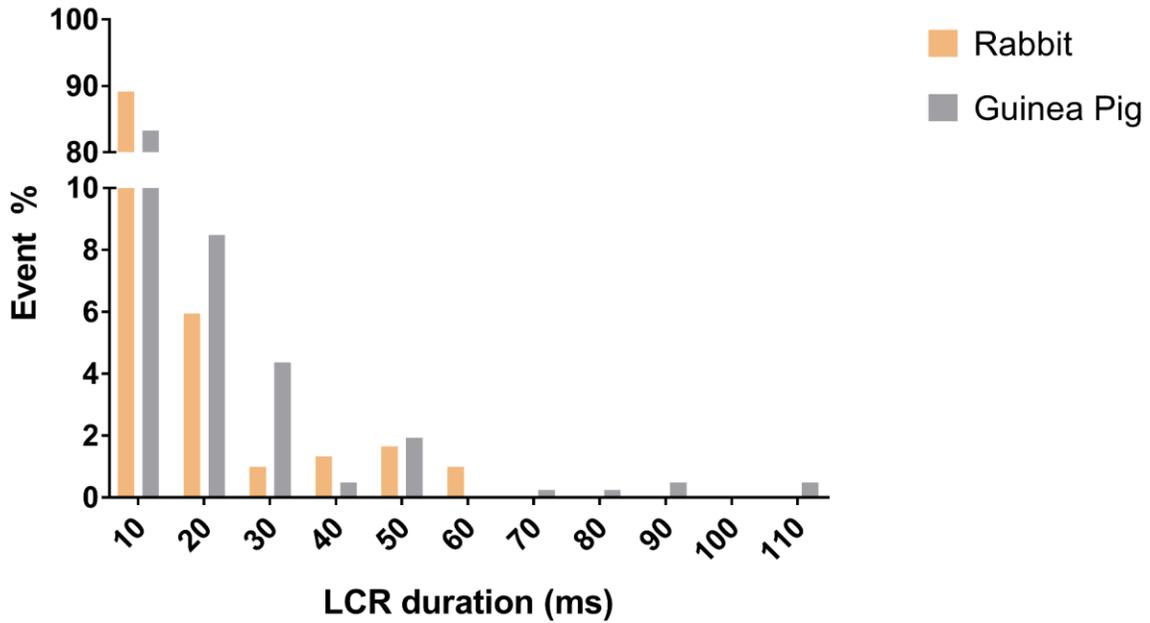

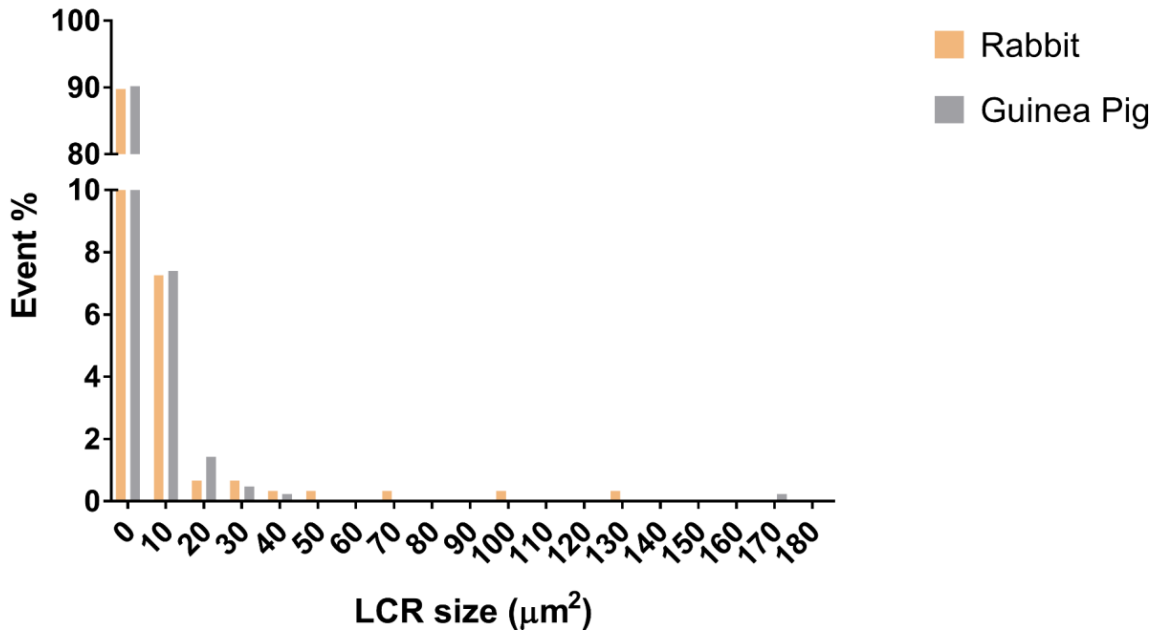

**Fig 6. Histograms showing distribution of LCR duration and size identified by XYTEventDetector.**
Histograms show substantial contribution of small, short-lived LCRs in both rabbit and guinea pig SAN. Data used in histograms are those presented in S1 Table 1. The LCR size was evaluated by output parameter Max Area scaled to µm$^2$.

*Noise considerations*

Any program detecting events from noise balances between two extremes: 1) Detecting events for sure, i.e. with no false-positive events, but inevitably missing important events closer to the noise; 2) Detecting almost all events, including, small ones, but inevitably catching many false-positive events. Varying the parameters defining sensitivity of LCR detection yields a variety of outputs for LCRs, including the aforementioned extreme cases. Our program is a tool to detect LCRs, rather than a definitive detection algorithm covering all cells and all possible recording conditions. This level of flexibility enables researchers to tune the sensitivity of detection in their studies. Depending on the specific hypothesis to be tested, this detection sensitivity (and the rate of false positive events) can be shifted. For example, if a study is focused on small events, close to the noise, then a larger (but still small) fraction of false positive events is inevitable and must be evaluated and care taken by statistical methods to test the hypothesis. If a study focuses on larger events, then the false-free detection can be substantially improved by decreasing sensitivity level far from the recording noise.

To ensure that we find real LCRs rather than noise, we developed a test to evaluate the rate of false-positive events, which can be used to tune the program for the optimal performance (illustrated in Fig 7). We cut a "control" sub-stack of images within the original data but far from the cell area. This control stack has instrumental noise, but no LCRs. Then we run the same LCR detection algorithm in both the control stack and in the cell area. The program reports the cell area in pixels, making it easy to normalize the number of false positive events found in the control stack area to all events found in the cell area. Our representative measurements of LCR parameters in S1 Table are provided along with an indication of false positive rates in each measurement. With the specific detection parameters, we were able to keep the rate of false positive events low, 0.97% and 6.7% in two rabbit cells and 0% in 6 other cells tested. Thus, this test indicates that the program parameters can be tuned to detect with a high probability true LCR events, rather than noise.

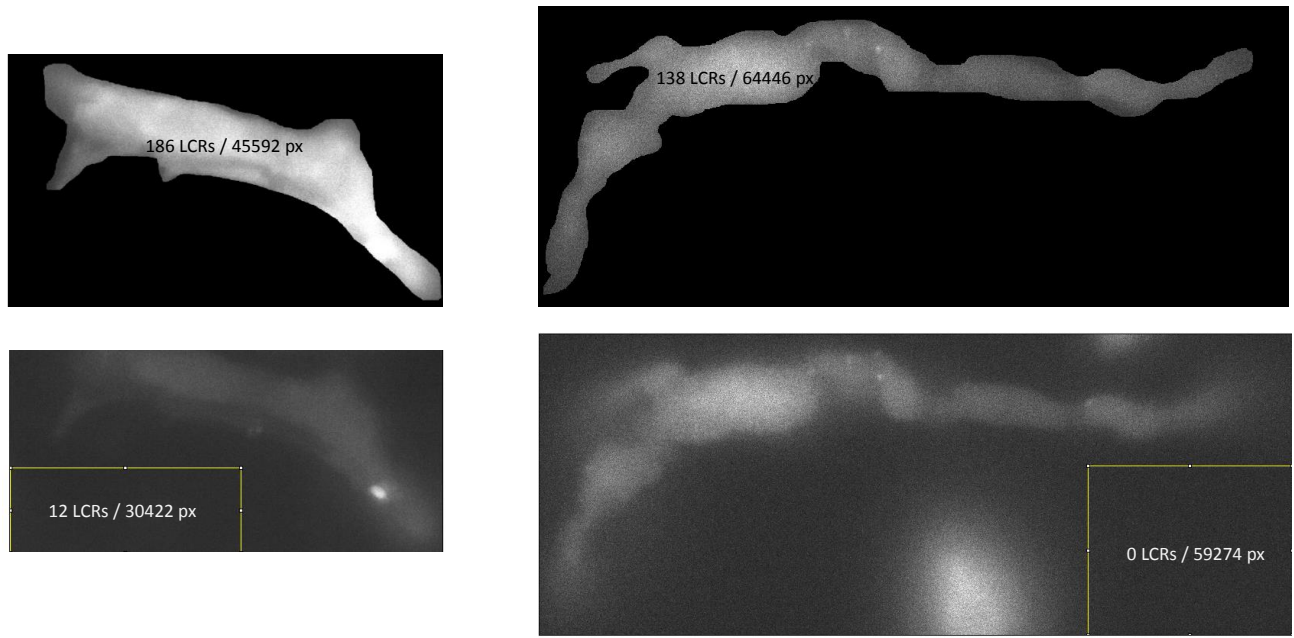

**Fig 7. Examples of the evaluation of the rate of false-positive events.**
Examples are given for SAN from two species, rabbit on the left and guinea pig on the right. The LCRs were found first within the cell area and then using the same procedure in the recording area marked by the yellow square, far from the cell. The number of detected LCRs are shown together with total number of pixels of respective search areas.

It is important to note that our program represents an exploratory tool to detect $Ca^{2+}$ release events (or any signaling events) from noise in 2D recording. There is no definitive or universal set of detection parameters that provides immediate and optimized LCR detection, because of the heterogeneity of cells, indicator loading (experiencing then various rates of bleaching and leaking), indicator excitation level, $Ca^{2+}$ release flux, spatial and temporal resolution, quality of optics, camera noise, and other recording condition variations. A good practical marker is that LCRs should be visible by eye in the movie after filtering and contrasting to ensure that the program will work. It is desirable to have a camera spatial resolution that is close to or surpass the optical limit for a light microscope which is around 250 nm.

**Discussion**

Here, we have developed a novel and unique set of computer algorithms that allow the automatic detection and analysis of LCRs in spontaneously contracting SA node cells. The capabilities of the programs can be potentially applied to any kind of a spontaneously contracting cell. To achieve this outcome, it has been necessary to overcome many technical challenges, including cell contraction motion artifacts and recording noise. We have also developed new terminology and algorithms for describing LCR path area, collisions, and separations, which represent important, previously undescribed characteristics of LCR behavior.

The entire set of algorithms was developed de novo, and has now been made available for further evaluation, development and customization to test particular hypotheses within the boundaries of specific recording conditions (resolution and signal to noise ratio).

The importance of these new tools for future research in local $Ca^{2+}$ events is that they allow rapid, objective, detailed, effective and unbiased detection of LCR events, thus providing new insights into their emergence and biophysical behavior. Although the focus of this paper has been on the development of these intricate methods, we have also clearly demonstrated several important fundamental characteristics of LCRs in SA node cells. Our results show that, in contrast to stereotyped $Ca^{2+}$ sparks in ventricular myocytes, the automatically detected LCRs are indeed extremely complex heterogeneous spatiotemporal events that propagate short distances, and often undergo mergers and separations (Fig 4).

Future studies using automatic LCR detection will identify the prevailing physiological conditions favoring LCRs being small, non-propagating spark-like events, versus large, complex widely-propagating events. Based on prior studies using confocal microscopy, the smaller events are expected to occur when cells decrease their pacemaker rate under cholinergic receptor (parasympathetic) stimulation [17], whereas larger and more synchronized events are expected when cells increase their rate in the presence of beta-adrenergic receptor (sympathetic) stimulation [18][19]. Thus, our new algorithms will be important in uncovering the intricate and detailed mechanisms of autonomic nervous system control of automaticity via LCRs and phosphorylation [20]. Previous studies of LCRs have been incomplete and imprecise due to necessary manual detection and analysis of LCRs along confocal scan lines, or in 2D movies. Our quick test of the new programs detailed above has already revealed substantially more LCRs per cycle, including a subtype of small LCRs (Fig 6), which are hard to reliably resolve manually without clear objective criteria. This presence of small, short lived LCRs has been recently predicted by numerical modeling of $Ca^{2+}$ releases generated by RyR clusters of different sizes (including very small clusters) identified by immunofluorescence confocal studies in 3D [21].

The much larger numbers of automatically detected LCR events enabled by the new programs will facilitate highly rigorous statistical analysis of LCR characteristics, which is required to study their stochastic nature in fine detail, and also to link their complex spatiotemporal parameters to the hierarchical structure of release channel distribution found in pacemaker cells [21]. Another important future study will involve comparison of the experimentally detected LCRs with numerically simulated LCRs [21, 22], utilizing the same automatic LCR detection programs in both cases. This will validate the recent numerical models of pacemaker cells featuring local Ca$^{2+}$ control [21].

The algorithms developed here are also expected to be helpful in studies of physiology and biophysics of Ca$^{2+}$ signaling in other cell types, including neurons [23] and muscle cells of non-cardiac origin, in which Ca$^{2+}$ signals are typically challenging to detect due to motion artifacts. Indeed, the diverse gamut of LCRs, Ca$^{2+}$ wavelets, and abortive Ca$^{2+}$ waves demonstrated here in SA node cells also emerge in ventricular myocytes [24] and Purkinje cells [25] under different experimental and physiological conditions, and would be also characterized by the new programs. Specifically, phosphorylation of sarcoplasmic reticulum Ca$^{2+}$ cycling proteins unleashes a high-power, rhythmical LCRs in ventricular myocytes linked to arrhythmias and relevant to biological pacemaker design [24], and would benefit from full characterization using our novel programs.

*Study Limitations*

SANC Analyser relies on the cell being orientated in the horizontal plane for its movement subtraction capabilities to be fully realized. It also relies on the cell being studied not undergoing any translational movement during the time of study, i.e. that cells being studied are well adhered to the dish in which they are being studied. Some SAN cells in particular are highly complex morphologically and have sea-weed like fronds that do not always adhere to the study dish, and as such LCR events in these complex appendages are not possible to quantify using the current program. SANC Analyser however does not rely on cells being long and thin like a typical SAN cell – more uniform shapes such as those typically seen with rounded stem cell-derived cardiomyocytes would also be appropriately handled by the program.

XYTEventDetector detected substantial number of small, short-lived LCRs (Fig 6). Due to the 10 ms sampling interval of our recording cameras, the duration of many small LCRs is 1 sample. While detecting these small LCRs is important in studies of biophysical mechanisms of SANC function, the exact shape and duration of these 1 sample events is imprecise and these data should be treated and interpreted with caution.

It is important to note that increasing just sampling rate of 2D imaging may not necessarily provide a better LCR detection and time profiling. In the present study, we also tested LCR recording with 5 times higher sampling rate (500 vs. 100 frames/s) using a modern PCO.edge 4.2 CMOS camera. However, many LCRs were not detected by our program at this extremely fast sampling rate because to catch LCRs with same success, the signal/noise should also be improved by a factor of 5. Thus, the most reliable detection requires that sampling interval should be close to the characteristic time for signal rise to peak, typically 5-10 ms for sparks and LCRs. However, we want to emphasize that since our algorithms reliably detect and accurately evaluate durations of longer events, this 10 ms limitation is due to temporary constrains of current technologies in recording stronger intracellular calcium signals at higher rates in 2 dimensions, rather than fundamental limits of our algorithms.

Data presented in S1 and S2 Tables were collected form a few cells in a few cycles to illustrate the function of our algorithms and typical parameters of our analysis. The values of specific LCR characteristics in the tables vary greatly and should be interpreted with cation.

Future extended studies are needed for further, more precise evaluation of LCRs in 2 dimensions.

*Conclusions*

Our study describes in detail novel, specific algorithms for automatic detection and analysis of LCRs occurring in 2D recordings in live, spontaneously beating pacemaker cells. One algorithm tracks points along the midline of the moving/contracting cell, using these points as a coordinate system for affine transform, producing a transformed image series where the cell does not move/contract. Thereafter, a second algorithm produces detailed descriptions of major LCR parameters such as period, signal mass, duration, and propagation path. As the LCRs propagate in live cells, the algorithm identifies complex and elegant splitting and merging behaviors, indicating the importance of locally propagating CICR for the fate of LCRs, and building up a powerful ensemble $Ca^{2+}$ signal. While the algorithms were developed to detect LCRs in sinoatrial nodal cells, they have the potential to be used in other applications in biophysics and cell physiology, for example, to detect $Ca^{2+}$ wavelets (abortive waves), sparks and embers in muscle cells [2] and $Ca^{2+}$ puffs and syntillas in neurons [16].

*Funding Statement*

The work was supported by the Intramural Research Program of the National Institute on Aging, National Institute of Health. SP was supported the Canadian Institutes of Health Research (MOP12874) and the Natural Sciences and Engineering Research Council (386877). The funders had no role in study design, data collection and analysis, decision to publish, or preparation of the manuscript.

10. Torrente AG, Mesirca P, Neco P, Rizzetto R, Dubel S, Barrere C, et al. L-type Cav1.3 channels regulate ryanodine receptor-dependent Ca2+ release during sino-atrial node pacemaker activity. Cardiovasc Res. 2016;109(3):451-61.

11. Lakatta EG, Maltsev VA, Vinogradova TM. A coupled SYSTEM of intracellular $Ca^{2+}$ clocks and surface membrane voltage clocks controls the timekeeping mechanism of the heart's pacemaker. Circ Res. 2010;106:659-73.

12. Monfredi O, Maltseva LA, Spurgeon HA, Boyett MR, Lakatta EG, Maltsev VA. Beat-to-beat variation in periodicity of local calcium releases contributes to intrinsic variations of spontaneous cycle length in isolated single sinoatrial node cells. PLoS One. 2013;8(6):e67247.

13. Vinogradova TM, Zhou YY, Bogdanov KY, Yang D, Kuschel M, Cheng H, et al. Sinoatrial node pacemaker activity requires $Ca^{2+}$/calmodulin-dependent protein kinase II activation. Circ Res. 2000;87(9):760-7.

14. Steinier J, Termonia Y, Deltour J. Smoothing and differentiation of data by simplified least square procedure. Anal Chem. 1972;44(11):1906-9.

15. Maltsev VA, Lakatta EG. Synergism of coupled subsarcolemmal $Ca^{2+}$ clocks and sarcolemmal voltage clocks confers robust and flexible pacemaker function in a novel pacemaker cell model. Am J Physiol Heart Circ Physiol. 2009;296(3):H594-H615.

16. Berridge MJ. Calcium microdomains: organization and function. Cell Calcium. 2006;40(5-6):405-12.

17. Lyashkov AE, Vinogradova TM, Zahanich I, Li Y, Younes A, Nuss HB, et al. Cholinergic receptor signaling modulates spontaneous firing of sinoatrial nodal cells via integrated effects35

# Supporting information

**Supplemental Tables**

**S1 Table.** Results of LCR detection analysis performed by XYTEventDetector using Ca signal recordings by **PCO.edge 4.2 CMOS camera** in SA node cells of rabbit and guinea pig, 4 cells of each species.

| Parameters | Rabbit | | | | | Guinea Pig | | | | |
|---|---|---|---|---|---|---|---|---|---|---|
| | cell#1 | cell#2 | cell#3 | cell#4 | Mean | cell#1 | cell#2 | cell#3 | cell#4 | Mean |
| Cycle Length (ms) | 540 | 690 | 480 | 560 | 567.5 | 1370 | 690 | 380 | 800 | 810.0 |
| Path Size ($\mu m^2$) | 2.427 | 5.211 | 3.01994 | 1.216 | 3.0 | 5.641 | 1.218 | 7.233 | 1.03 | 3.8 |
| Duration (ms) | 12.5 | 14.3 | 12.3 | 11.2 | 12.6 | 24.4 | 10.8 | 15.5 | 11.1 | 15.5 |
| LCR Period (ms) | 471.5 | 446.7 | 221.1 | 370.6 | 377.5 | 843 | 451.4 | 229.8 | 474.7 | 499.7 |
| LCRs/Cycle | 59 | 48 | 81 | 119 | 76.8 | 184 | 87 | 52 | 75 | 99.5 |
| False positives rate% | 0 | 0 | 6.7 | 0.97 | 1.9 | 0 | 0 | 0 | 0 | 0.0 |
| Max Filter | 100 | 100 | 75 | 100 | 93.8 | 100 | 100 | 100 | 100 | 100.0 |
| SD Detection | 0.4 | 0.4 | 0.4 | 0.4 | 0.4 | 0.4 | 0.3 | 0.3 | 0.3 | 0.3 |
| SD Termination | 1.5 | 1.5 | 1.5 | 1.5 | 1.5 | 1.5 | 1.5 | 1.5 | 1.5 | 1.5 |
| Search Distance | 8 | 8 | 8 | 4 | 7.0 | 8 | 7 | 8 | 7 | 7.5 |
| Size Threshold | 7 | 7 | 7 | 7 | 7.0 | 7 | 7 | 7 | 7 | 7.0 |
| Intensity Threshold | 50 | 50 | 50 | 50 | 50.0 | 50 | 50 | 50 | 50 | 50.0 |
| % Transient Cutoff | 99 | 70 | 90 | 90 | 87.3 | 70 | 70 | 70 | 70 | 70.0 |
| Cell size (µm2) | 683.13 | 281.73 | 654.91 | 415.91 | 508.9 | 483.28 | 925.5 | 1093.06 | 705.81 | 801.9 |

Results are mean values obtained in one pacemaker cycle in each cell. Grey area shows specific input program parameters to analyze recordings in each cell.

**S2 Table.** Results of LCR detection analysis performed by XYTEventDetector using Ca signal recordings by **Hamamatsu C9100-12 CCD camera** in SA node cells of rabbit and guinea pig, 4 cells of each species.

| Parameters | Rabbit | | | | | Guinea Pig | | | | |
|---|---|---|---|---|---|---|---|---|---|---|
| | cell#1 | cell#2 | cell#3 | cell#4 | Mean | cell#1 | cell#2 | cell#3 | cell#4 | Mean |
| Cycle Length (ms) | 536 | 480 | 361 | 512 | 472.25 | 577 | 515 | 457 | 411 | 490 |
| Path Size ($\mu m^2$) | 9.56 | 8.96 | 10.44 | 6.75 | 8.92 | 9.88 | 12.43 | 11.38 | 10.18 | 11.0 |
| Duration (ms) | 14.29 | 13.2 | 14.32 | 14.18 | 14.0 | 13.45 | 16.33 | 15.75 | 13.52 | 14.76 |
| LCR Period (ms) | 230.27 | 209.85 | 126.97 | 193.1 | 190.0 | 369.57 | 269.18 | 260.09 | 257.32 | 289.04 |
| LCRs/Cycle | 61 | 66 | 44 | 98 | 67.25 | 86 | 53 | 38 | 47 | 56 |
| False positives rate% | N/D | N/D | N/D | N/D | | N/D | N/D | N/D | N/D | |
| Max Filter | 80 | 80 | 170 | 150 | | 100 | 150 | 100 | 100 | 112.5 |
| SD Detection | 0.4 | 0.5 | 0.5 | 0.5 | | 0.3 | 0.4 | 0.4 | 0.4 | 0.375 |
| SD Termination | 1 | 2 | 1 | 1 | | 1 | 1 | 1 | 1 | 1 |
| Search Distance | 3 | 3 | 3 | 3 | | 4 | 4 | 4 | 4 | 4 |
| Size Threshold | 5 | 5 | 5 | 5 | | 5 | 5 | 5 | 5 | 5 |
| Intensity Threshold | 50 | 50 | 50 | 50 | | 50 | 50 | 50 | 50 | 50 |
| % Transient Cutoff | 99 | 98 | 99 | 99 | | 99 | 99 | 99 | 99 | |
| Cell size (µm2) | N/D | N/D | N/D | N/D | | N/D | N/D | N/D | N/D | 99 |

Results are mean values obtained in 3 pacemaker cycles in each cell. N/D, not determined. Grey area shows specific input program parameters to analyze recordings in each cell.

**Supplemental Movies**

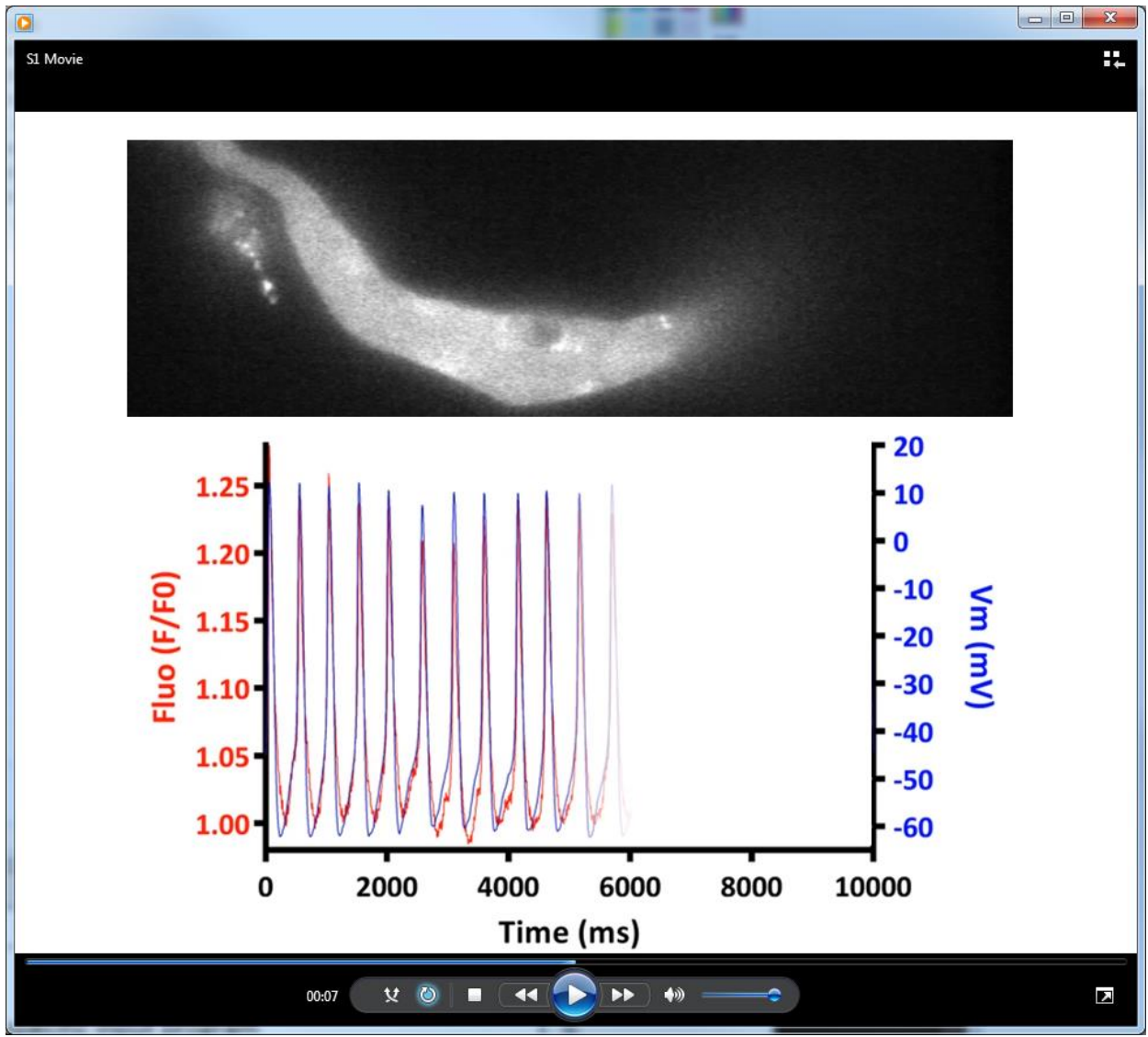

*Click image for web Link to S1 Movie.mp4*

**S1 Movie.** An example of simultaneous recording of intracellular $Ca^{2+}$ (fluo-4) and membrane potential (perforated patch clamp). LCRs occur between AP-induced $Ca^{2+}$ transients. The $Ca^{2+}$ signal in this movie is taken from the whole cell area.

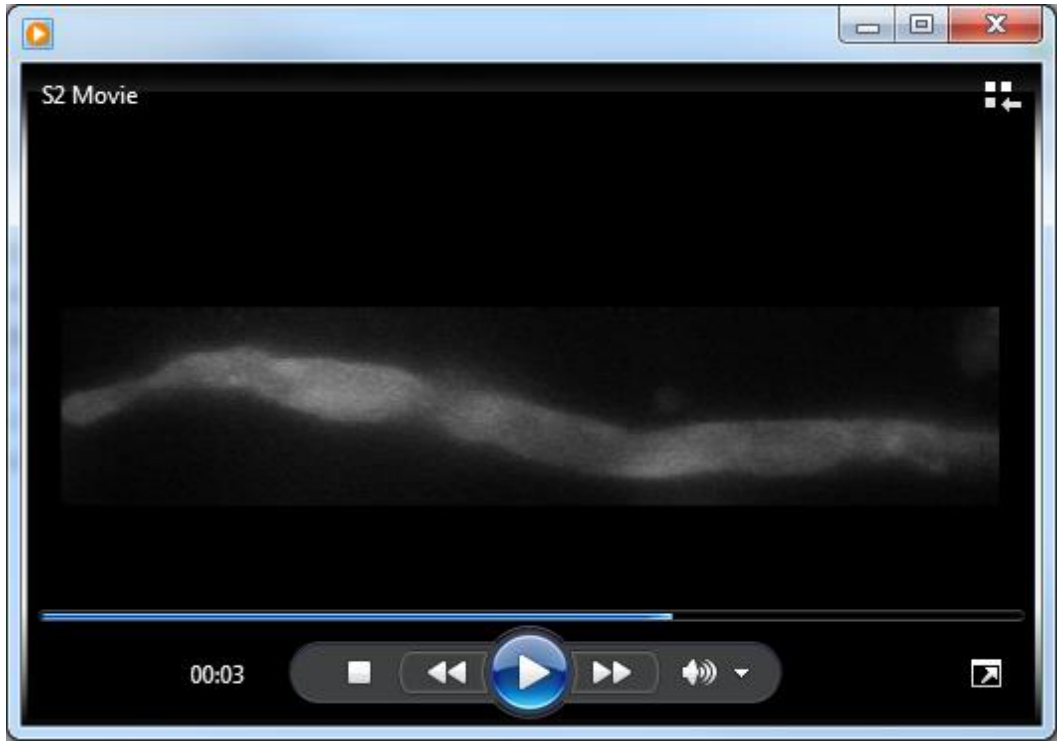

[Click image for web Link to S2 Movie.wmv](#)

**S2 Movie.** An example of original recording of intracellular $Ca^{2+}$ signals in spontaneously contracting SA node cell before application of SANC Analyser. The $Ca^{2+}$ imaging data were acquired by Hamamatsu C9100-12 CCD camera and $Ca^{2+}$ indicator fluo-4. The duration of the recording is 600 ms.

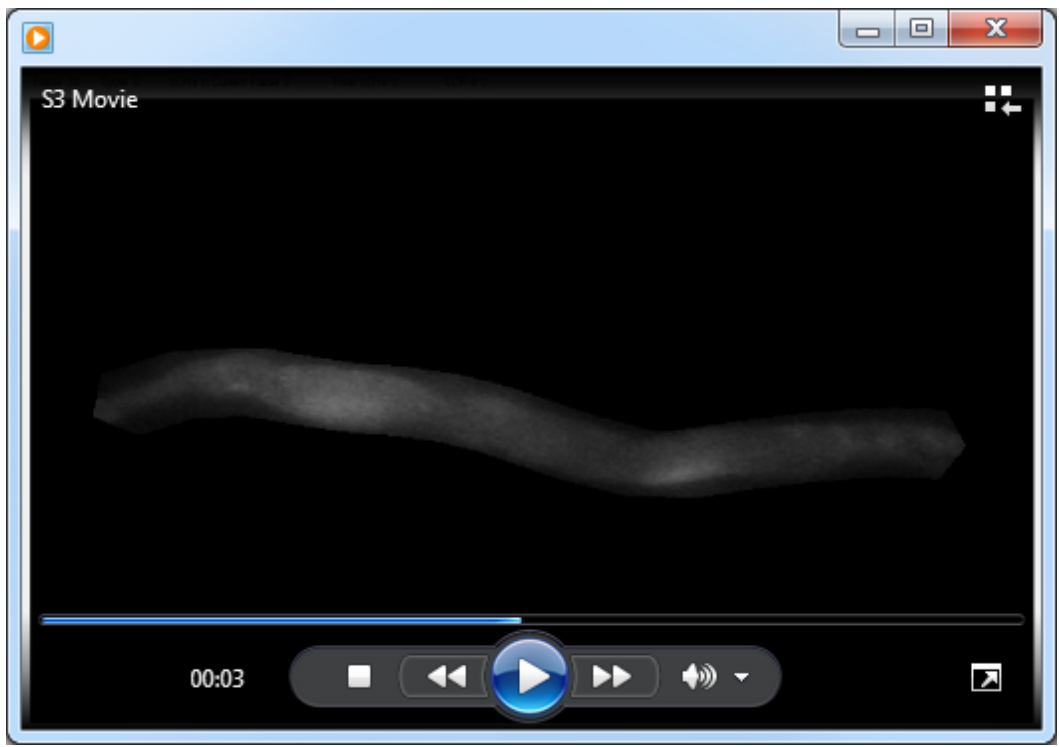

*Click image for web Link to S3 Movie.wmv*

**S3 Movie.** The same cell as in S2 Movie but after application SANC Analyser that tracks contractile motion and make cell to appears as motionless but with all local signalling preserves assigning each given cytoplasm location to each specific pixel in the movie. The duration of the recording is 600 ms.

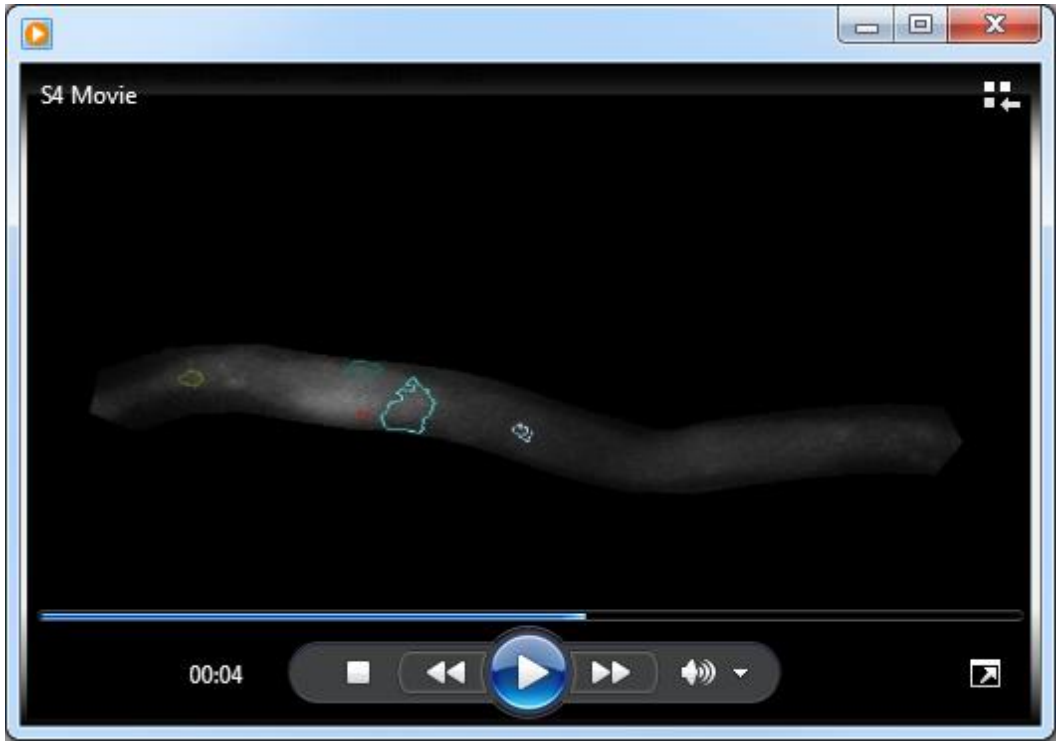

[Click image for web Link to S4 Movie.wmv](#)

**S4 Movie.** An example of LCRs identified by XYTEventDetector program. Perimeter of each detected LCR is outlined by a color border for clear visualization. The duration of the recording is 600 ms.

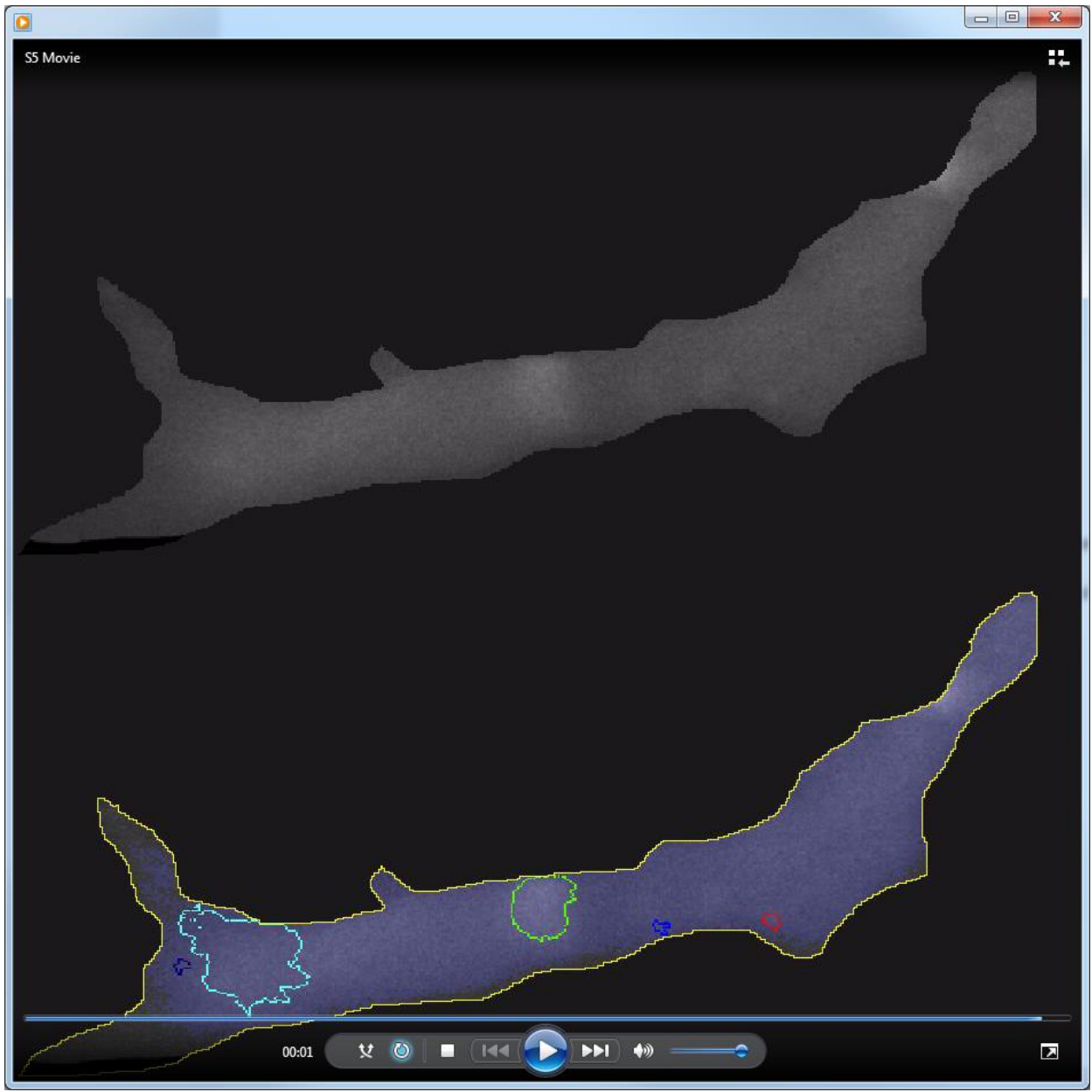

[Click image for web Link to S5 Movie.avi](#)

**S5 Movie.** An example of LCRs identified by XYTEventDetector program in a guinea pig SANC with PCO.edge 4.2 CMOS camera. Perimeter of each detected LCR is outlined by a color border for clear visualization. The duration of the recording is 1370 ms (Guinea pig cell #1 in S1 Table).